\documentstyle{elsart}

\begin{document}

\begin{frontmatter}

\title{Role of Group and Phase Velocity in High-Energy Neutrino Observatories}
\author{P.\ B.\ Price and K.\ Woschnagg}
\address{Physics Department, University of California, Berkeley, CA 94720, USA}

\begin{abstract}
Kuzmichev recently showed that use of phase velocity rather than group
velocity for Cherenkov light signals and pulses from calibration lasers
in high-energy neutrino telescopes leads to errors in track reconstruction
and distance measurement.
We amplify on his remarks and show that errors for four cases of interest
to AMANDA, IceCube, and RICE (radio Cherenkov detector) are negligibly small.
\end{abstract}

\end{frontmatter}

\section{Introduction}

Kuzmichev~\cite{Kuzmichev} pointed out that participants in high-energy
neutrino projects mistakenly use tabulated values of refractive index,
$n$, which corresponds to phase velocity, instead of using values of
group refractive index, $n_g$, which applies to signals transmitted at
group velocity, $v_g$, where $n_g(\omega)\equiv c/v_g(\omega)$.
The choice of $n$ rather than $n_g$ or of $v$ rather than $v_g$ leads to
errors in track reconstruction and in distance measurements.

The subject of which velocity to use for Cherenkov light has an interesting
history. Tamm~\cite{Tamm} was the first to point out (correctly) that for
a particle moving with speed $\beta$c the wavefront of a Cherenkov
``bow wave'' is a cone (the ``Mach cone'') with semi-angle
\begin{equation}
\cot\alpha = (\beta^2n^2-1)^{1/2} 
           + n\omega\beta^2\frac{{\rm d}n}{{\rm d}\omega}/
                  (\beta^2n^2-1)^{1/2}
\label{semiangle}
\end{equation}
whereas the Cherenkov radiation is emitted at the angle $\theta$ given by
\begin{equation}
\cos\theta = \frac{1}{\beta n}
\end{equation}
The presence of the second term in eq.(\ref{semiangle}) results in $\theta$
not being the complement of $\alpha$ and in the direction of light emission
not being exactly normal to the wavefront. Unfortunately, most textbooks
draw ray vectors normal to the bow wave. Surprisingly, even 
Sommerfeld~\cite{Sommerfeld} wrote, ``The ray direction is perpendicular
to the Mach cone.''

Motz and Schiff~\cite{Motz} corrected the Sommerfeld statement, showing
clearly that ``the radiation propagates in directions that make the angle
$\theta$ with the path of the [moving] charge, where [$\cos\theta=1/\beta n$],
so that the cone may be thought of as `side-slipping' as it moves along
with the charge.'' They derived eq.(\ref{semiangle}) and showed that only
if ${\rm d}n/{\rm d}\omega=0$ would the ray direction be normal to the cone.

We now discuss several ways in which data for AMANDA and other neutrino
telescopes need to be corrected, and show that the corrections are mostly
very small.

\section{Direction and velocity of Cherenkov light from a muon in ice or water}

The direction of propagation of the Cherenkov light does not need correction,
but the velocity does.
Using an extensive table of data~\cite{Warren} and interpolating linearly in
$\log\lambda$ for phase refractive index for ice, we generated a polynomial
(with wavelength $\lambda$ in micrometers)
\begin{equation}
n(\lambda) = 1.55749 - 1.57988\lambda + 3.99993\lambda^2 - 4.68271\lambda^3
           + 2.09354\lambda^4
\label{neq}
\end{equation}
in the region 0.3 $\mu$m to 0.6 $\mu$m for which Cherenkov light 
contributes to the signal of a muon.
For group refractive index, we generated a polynomial for the fractional
increase
\begin{equation}
\frac{n_g-n}{n} = \frac{\lambda}{n}\left|\frac{{\rm d}n}{{\rm d}\lambda}\right|
        = 0.227106 - 0.954648\lambda + 1.42568\lambda^2 - 0.711832\lambda^3
\end{equation}
where ${\rm d}n/{\rm d}\lambda$ is obtained by differentiating (\ref{neq}).
The required correction for distances traveled by Cherenkov light from 
points on a muon trajectory to phototubes decreases monotonically from
-5.0\% at the shortest wavelength, 0.3~$\mu$m, to -1.4\% at the longest
wavelength, 0.6~$\mu$m, if $n(\lambda)$ had been used in previous analyses.
However, the AMANDA collaboration simply used a fixed value, $n=1.32$,
for all wavelengths.
The correction to this value would then range from -6.1\% at 0.3~$\mu$m to
-0.5\% at 0.6~$\mu$m.
In all cases the true distance is less than the value previously used.
Taking into account the Cherenkov spectrum, the transparency of the glass
pressure vessel, and the phototube response, the most effective wavelength
is around 0.38 $\mu$m, with a corresponding correction of -3\% for distances.
Phototube jitter is $\sim$10 ns.
The strongest cut for AMANDA track reconstruction is that $N_{{\rm dir}}>4$,
where $N_{{\rm dir}}$ is the number of ``direct hits,'' defined as Cherenkov
photon arrival times between -15 and +50 ns of those expected for no scattering.
A 3\% error in distance corresponds to a 2 ns error in timing,
which is much smaller than phototube jitter.

\section{Use of pulsed lasers to measure spacing between strings}

Since distance is inferred from arrival time of the leading edge of a pulse
distribution, distance is proportional to pulse velocity and inversely
proportional to refractive index.
All horizontal distances within the AMANDA detector, i.e.\ the interstring
separations, must be reduced by the factor $n/n_g$.
A YAG laser with wavelength 0.532 $\mu$m was used for distance measurements.
Distances must therefore be reduced by the factor $1.32/n_g(0.532)=0.99$.

\section{Use of pulsed light sources to infer wavelength-dependent absorption
length, $\lambda_a$, and wavelength-dependent effective scattering length,
$\lambda_e$}

For distances $d\gg\lambda_e\approx 25$ m, the expression for three-dimensional
random walk with absorption gives an excellent approximation to the propagation
of light pulses in ice~\cite{Askebjer}:
\begin{equation}
u(d,t) = \left(\frac{3}{4\pi v_g\lambda_et}\right)^{3/2}
\exp{\left\{-\frac{3d^2}{4v_g\lambda_et}-\frac{v_gt}{\lambda_a}\right\}}
\label{diffeq}
\end{equation}
Ignoring the pre-exponential factor, we see that the estimate of $\lambda_a$
scales as $n_g$ and the estimate of $\lambda_e$ scales as $d^2/n_g$.
If distance had been measured with a meter stick or other method that did
not involve light pulses, the estimate of $\lambda_e$ would scale only as
$1/n_g$. However, we measured distances with light pulses, as a consequence
of which the net scaling should be as $n_g$. As a specific example, for two
strings $\geq 100$ m apart, eq.(\ref{diffeq}) is a good approximation,
and we find that both $\lambda_a$ and $\lambda_e$ need to be decreased
by 1\%, for studies carried out with a YAG laser. As an independent check,
we generated a function $u(d,t)$ for values of $\lambda_a=20$ to $100$ m,
$\lambda_e=15$ to $45$ m, and distances 100 to 300 m. We first used the
phase refractive index to generate $u(d,t)$; then we increased $n$ by
fractional amounts ranging from 0.5\% to 5\% and evaluated the new values
of $\lambda_a$ and $\lambda_e$ that gave the best fit. For separations
inferred from the pulse method, we found that $\delta\lambda/\lambda$
increases as $(\delta n/n)^{\kappa}$ in both cases, with 
$\kappa=1.03\pm 0.02$ for $\lambda_a$, 
and $\kappa=1.01\pm 0.02$ for $\lambda_e$.
For separations assumed to have been measured with a meter stick, the
best-fit powers were $\kappa=1.03\pm 0.02$ for $\lambda_a$ and
$\kappa=-1.01\pm 0.02$ for $\lambda_e$.
Thus, both methods lead to the same conclusion regarding scaling of absorption
and scattering lengths.

\section{Use of pulses for timing calibration}

In AMANDA, the individual time offsets ($t_0$) between the times recorded
by the surface electronics and the times when photons hit the phototubes
are measured with YAG laser pulses (0.532 $\mu$m) sent down optical fibers
to diffusive balls near or inside the phototubes. For most phototubes
a fiberball located within 1 m is used, but for 15\% of the phototubes
the pulses have to travel through at least 10 m of ice. The time for this
propagation through ice has to be corrected by $n_g/n=1.01$, which for
distances of 10 to 30 m corresponds to 0.44 to 1.3 ns. This correction
is smaller than the overall measurement error of individual timing offsets
and is considerably smaller than the uncertainty in the recorded times of
light pulses.

\section{Dependence of group refractive index on density and temperature}

AMANDA-B is instrumented at depths from 1.5 to 1.9 km, and IceCube will
be instrumented at depths from 1.4 to 2.4 km.
The change in refractive index with density is given by
\begin{equation}
\frac{{\rm d}n}{{\rm d}\rho} = \frac{(1-\lambda_0)(n^2+2)(n^2-1)}{6n\rho}
\end{equation}
where $\lambda_0$ is a small correction due to strain-induced 
polarizability~\cite{Burstein}.
Over a depth from 1.4 to 2.4 km, the density of ice actually decreases
slightly, due to the increase in temperature with depth.
From data of Gow~\cite{Petrenko}, we estimate 
$\delta\rho/\rho\approx 0.004$ over
a 1 km change of depth from 1.4 to 2.4 km, which leads to 
$\delta n/n\approx 0.002$.
Over a depth from 2.0 to 2.3 km in seawater (applicable to ANTARES),
$\delta n/n\approx 0.0004$ due to the density increase.

For IceCube, the ambient ice temperature increases from -41$^\circ$C to 
-19$^\circ$C
for a change of depth from 1.4 to 2.4 km.
The measured change in refractive index of ice with temperature is 
given~\cite{Petrenko} by
\begin{equation}
\frac{{\rm d}n}{{\rm d}T} = -2.7\times 10^{-5} n
\end{equation}
which leads to a fractional decrease $\delta n/n=-6\times 10^{-4}$ for
the 1 km increase in depth.
In Mediterranean seawater the temperature decreases about 1$^\circ$ per km at
depths relevant to ANTARES; the resulting change of refractive index due
to a change of temperature is negligibly small.

\section{Conclusions}

For 0.532 $\mu$m pulses in ice, no error exceeds $\sim$1\% when a single value
of 1.32 for refractive index was previously used.
If a nitrogen laser (0.337 $\mu$m) were to be used, the error would grow
to 5\% relative to the value 1.32.

For radio Cherenkov radiation in ice (RICE), ${\rm d}n/{\rm d}\lambda=0$ for
wavelengths from a few cm to $\sim$10 m, 
which spans the entire region of interest.
Thus, values of the phase refractive index can be used without introducing
error in analysis of RICE data.

For ANTARES and other neutrino telescopes in seawater, 
$\delta n/n\approx 0.0012$ over a depth interval of 1 km, due to the
dependence of $n$ on pressure; the change due to dependence of $n$ on
temperature is negligible.

For AMANDA and IceCube, the change in refractive index due to the increase
of ice density leads to an error of only $\sim$0.2\% per km.
The error due to a temperature increase with depth would be even smaller,
$\sim$0.06\%.

\section*{Acknowledgments}

This research was supported in part by National Science Foundation grant
PHY-99-71390.


\begin{thebibliography}{100}

\bibitem{Kuzmichev}
{L.\ A.\ Kuzmichev, hep-ex/0005036, 25 May 2000.}

\bibitem{Tamm}
{I.\ Tamm, J.\ Phys.\ U.S.S.R.\ {\bf 1}, 439 (1939).}

\bibitem{Sommerfeld}
{A.\ Sommerfeld, {\em Vorlesungen \"{u}ber Theoretische Physik}, Band IV,
Optik (Dieterich Verlag, Wiesbaden, 1950), p.\ 341.}

\bibitem{Motz}
{H.\ Motz and L.\ I.\ Schiff, Am.\ J.\ Phys.\ {\bf 21}, 258 (1953).}

\bibitem{Warren}
{S.\ G.\ Warren, Appl.\ Optics {\bf 23}, 1206 (1984).}

\bibitem{Askebjer}
{P.\ Askebjer {\it et al.}},
{Geophys.\ Res.\ Lett.\ {\bf 24}, 1355 (1997).}

\bibitem{Burstein}
{E.\ Burstein and P.\ L.\ Smith},
{Phys.\ Rev.\ {\bf 74}, 229 (1948)}.

\bibitem{Petrenko}
{V.\ F.\ Petrenko and R.\ W.\ Whitworth},
{\em Physics of Ice} (Oxford University Press, Oxford, 1999), p.\ 219.

\end{thebibliography}
\end{document}